\documentclass[letterpaper,twocolumn,10pt]{article}

\usepackage{usenix,epsfig, endnotes}

\usepackage{booktabs}
\usepackage{xspace}
\usepackage{graphicx}
\usepackage{rotating}
\usepackage{multirow,bigdelim}
\usepackage{comment}
\usepackage{pifont}
\usepackage{xcolor}
\usepackage[nointegrals]{wasysym}
\usepackage{tikz}
\usepackage{hyperref}

\definecolor{myred}{RGB}{255, 0, 0}
\definecolor{mycolor}{RGB}{0, 167, 159}
\definecolor{mygray}{RGB}{202, 199, 199}
\colorlet{light-gray}{mygray!50}
\colorlet{light-my}{mycolor!60}
\definecolor{lighter-gray}{gray}{0.97}

\newcommand{\kbox}[1]{
\smallskip
\begin{tikzpicture}

\node [fill=lighter-gray, text width=0.92\linewidth] (0,0) (mybox)
{ #1 };
\draw [mycolor, thick] (mybox.south west) -- (mybox.north west);
\draw [mycolor,thick] (mybox.south east) -- (mybox.north east);
\end{tikzpicture}}

\newcommand{\new}[1]{{\textcolor{black} {#1}}}

\newcommand{\no}{\ensuremath{\ocircle}\xspace}
\newcommand{\yes}{\ensuremath{\CIRCLE}\xspace}
\newcommand{\sometimes}{\LEFTcircle\xspace}

\newcommand{\myparagraph}[1]{\smallskip \noindent \textbf{#1.}}
\newcommand{\question}[1]{\ensuremath{\mathcal{Q}}{#1}}

\newcommand{\checkmark}{\ding{51}}

\newcommand{\classifier}{\ensuremath{\mathsf{F}}\xspace}
\newcommand{\params}{\ensuremath{\omega}\xspace}

\newcommand{\sample}{\ensuremath{\mathsf{x}}\xspace}
\newcommand{\lab}{\ensuremath{\mathsf{y}}\xspace}
\newcommand{\trainx}{\ensuremath{\mathsf{X}}\xspace}
\newcommand{\trainy}{\ensuremath{\mathsf{Y}}\xspace}
\newcommand{\testx}{\ensuremath{\mathsf{X_t}}\xspace}
\newcommand{\testy}{\ensuremath{\mathsf{Y_t}}\xspace}

\begin{document}
%

\date{}

\title{\Large \bf Towards More Practical Threat Models \\ in Artificial Intelligence Security}

\author{
{\rm Kathrin Grosse,$^1$ Lukas Bieringer,$^2$ Tarek R. Besold,$^3$ Alexandre Alahi$^1$}\\
$^1$EPFL, Switzerland, $^2$QuantPi, Germany, $^3$TU Eindhoven, The Netherlands
}


\maketitle

\thispagestyle{empty}

\begin{abstract}
Recent works have identified a gap between research and practice in artificial intelligence security: threats studied in academia do not always reflect the practical use and security risks of AI. For example, while models are often studied in isolation, they form part of larger ML pipelines in practice. Recent works also brought forward that adversarial manipulations introduced by academic attacks are impractical. 
We take a first step towards describing the full extent of this disparity. To this end, we 
 revisit the threat models of the six most studied attacks in AI security research and match them to AI usage in practice via a survey with \textbf{271} industrial practitioners. On the one hand, we find that all existing threat models are indeed applicable. On the other hand, there are significant mismatches: research is often too generous with the attacker, assuming access to information not frequently available in real-world settings. 
 Our paper is thus a call for action to study more practical threat models in artificial intelligence security. 
\end{abstract}


%

\section{Introduction}
A large body of academic work focuses on machine learning (ML) security~\cite{barreno2006can,biggio2018wild,chen2017targeted,cina2022wild,Dalvi:2004:AC:1014052.1014066,gu_badnets_2017,ji2017backdoor,oh2019towards,papernot2017practical,DBLP:journals/corr/SzegedyZSBEGF13,DBLP:conf/uss/TramerZJRR16}. 
Although these attacks have been established, increasing criticism targets their threat models. 
For example, most academic papers focus on standalone models~\cite{chen2017targeted,Dalvi:2004:AC:1014052.1014066,gu_badnets_2017,ji2017backdoor,oh2019towards,papernot2017practical,DBLP:journals/corr/SzegedyZSBEGF13,DBLP:conf/uss/TramerZJRR16}, while models in practice are generally embedded into pipelines or larger systems~\cite{evtimov2020security,bieringer2021mental}. 
In addition, it has been pointed out that attacks in practice do currently not require the degree of complexity inherent to academic publications~\cite{apruzzese2022real,grosse2022so}.
Also, the measurement of manipulations introduced by an attacker was deemed impractical~\cite{apruzzese2022real,gilmer2018motivating}, and the overall amount of data available to the attacker in some cases~\cite{cina2022wild,grosse2022so}.

For example, poisoning attacks~\cite{cina2022wild,rubinstein2009antidote,biggio2011support} require manipulating the training data. Grosse et al.~\cite{grosse2022so} reported cases of poisoning in the wild---yet it is unknown which fraction of companies allow access to their training data. Thus, the number of organizations vulnerable to poisoning attacks is, in practice, unknown. In addition, companies may only allow access to a fraction of their data---another limiting factor for an attack to succeed. As an example, consider a company where 1\% of the data can be accessed by the attacker. Most academic attacks require access to more data~\cite{cina2022wild}, limiting their usefulness. Analogously, evasion attacks were reported in the wild~\cite{grosse2022so}. Evasion requires the submission of at least one perturbed test sample~\cite{Dalvi:2004:AC:1014052.1014066,DBLP:journals/corr/SzegedyZSBEGF13}. Yet the number of AI systems in practice where this is possible is again unknown.

These works illustrate mismatches and demonstrate that some aspects of threat models are unaligned between research and practice. The underlying problem, an absence of knowledge on how artificial intelligence (AI) is used in practice, is however still unaddressed. In other words, it remains unknown whether researched threat models are \emph{representative} of AI usage in practice. We thus take a first step towards measuring this mismatch of AI security research and practice. 

\begin{table*}[ht]
\caption{Key Findings of our work.}\label{tab:findings}
\centering
\resizebox{\linewidth}{!}{\begin{tabular}{llr}
\toprule
& Key Finding (KF) & Section   \\ 
\midrule
KF 1 & \textbf{Access} is generally given to query the model and the corresponding outputs, or not at all. & \ref{sec:genthreatmodel} \\
KF 2 & \textbf{The underlying assumptions of all six attacks studied are relevant in practice.} & \ref{sec:specificAttacks} \\
KF 3 & Scientific threat models tend to be \textbf{too generous}: & \ref{sec:specificAttacks} \\
KF 3.1 & \hspace{1em} \textbf{Poisoning} and \textbf{backdoor} threat models assume unpractical fractions of alterable training data. & \ref{sec:trtimeatt} \\
KF 3.2 & \hspace{1em} \textbf{Black-box evasion} and \textbf{model stealing} threat models assume unpractical amounts of queries. & \ref{sec:testtime} \\
KF 3.3 & \hspace{1em} \textbf{Model stealing} and \textbf{privacy attacks} threat models' assumptions do not represent practical AI usage. & \ref{sec:testtime}\\
KF 4 & \textbf{Datasets} have often fewer features in practice than in AI security research. & \ref{sec:practChallenges} \\
KF 5 & \textbf{Code libraries} used in AI are security relevant. & \ref{sec:otherAttacks} \\
KF 6 & \textbf{AI security knowledge} does not influence the practical threat models of AI in our sample. & \ref{sec:threatmodelfactors} \\
\bottomrule
\end{tabular}}
\end{table*}

\textbf{Contributions.} To this end, we describe the commonly used academic threat models of the six most studied attacks in AI security: poisoning~\cite{cina2022wild}, backdoors~\cite{cina2022wild}, evasion or adversarial examples~\cite{Dalvi:2004:AC:1014052.1014066,DBLP:journals/corr/SzegedyZSBEGF13}, model stealing~\cite{DBLP:conf/uss/TramerZJRR16}, membership inference~\cite{shokri2017membership,choquette2021label}, and property inference~\cite{ateniese2015hacking,jegorova2022survey} in Section~\ref{sec:background}. To measure whether threat models match practical usage, we
 design a questionnaire in Section~\ref{sec:methodology} that collects information relevant to AI security like access patterns, data sources, etc. In the same section, we present our sample of \textbf{271} AI practitioners, before analyzing our results in Section~\ref{sec:results}. 

 We summarize our key findings in Table~\ref{tab:findings}. First, all six analyzed attacks are relevant in practice. 
In our sample, access to training data and the model is often constrained in practice, indicating overly generous assumptions in researched threat models. This includes large fractions of accessible training data for poisoning and backdoor attacks, and large query budgets for black-box evasion and model stealing. \new{Since} there are attacks with low budgets, \new{vulnerabilities can be exploited in practice and mitigations are needed.} Other mismatches between practice and research concern the used data, where academic datasets cover a large part of industrial datasets, but some cases are rarely studied. Finally, we aim to understand which factors influence threat models in practice. Here, knowledge of AI security has no influence. Only AI knowledge and AI maturity of the company are negatively correlated with public data sources, 
showing the need for future work.

We then revisit the limitations of our approach (Sect.~\ref{sec:limitations}). For example, our paper only provides initial insights about vulnerabilities via public access. Real-world threat surfaces may be larger. Still, the implications of our study (Sect.~\ref{sec:implications}) 
go beyond the above-discussed shortcomings of threat models. Implications also relate to current legislative attempts like the EU AI Act that requires security and vulnerability assessments of AI systems. We also set previously low numbers of AI security incidents into context and pave the way toward a deep understanding of what affects the security of AI-based products in practice. We then review related work (Sect.~\ref{sec:relwork}) and conclude our contributions (Sect.~\ref{sec:conclusion}).

\textbf{Remark.} \emph{This work should not be interpreted as a finger-pointing exercise. So far, AI security research has relied on best practices of security threat modeling, and we confirm that all 6 studied settings are applicable in practice. However, we describe unstudied settings hoping that we, as a community, can progress together toward more practical research.}

\section{Background}\label{sec:background}
Before we review AI threat models, we define  AI. To this end, we use the example of machine learning (ML), a sub-discipline of AI, and then outline the differences to other paradigms like reinforcement learning (RL) or data mining (DM). A typical task in machine learning is image recognition, e.g., classifying images from cats and dogs. In this case, we have a dataset of images \trainx and corresponding labels \trainy with the individual image \sample and label \lab. On this data, we train a classifier \classifier defined by its weights \params. We adjust these weights \params during training so that $\classifier(\params,\trainx)\approx\trainy$. The classifier \classifier then generalizes to unseen test images $\sample \in \testx$ and correctly predicts their labels $\lab \in \testy$.  
In contrast to the concrete label output used in this example, $\lab$ does not have to be discrete ('cat', 'dog') but can be continuous (regression) or more complex (in object detection or image segmentation).  
In the following, we refer to training data or \trainx, \trainy for any data used for model training. \new{Test data, or  \testx, refers to input during deployment (e.g. after the model development is complete). Test outputs, or \testy, to the corresponding outputs for a given \testx.}

RL, in contrast to ML, learns a policy that determines the behavior of an agent in an environment. Albeit different, also RL requires training and test data which can however take the form of an environment generating this data. DM analyses data and does not necessarily rely on test data. \new{The definitions of deployment data \testx or training data \trainx above are thus broad and encompass different formats like confidence scores and top-one outputs, as well as possible pre-processing. We skipped these details to encompass different paradigms and leave a detailed study of these aspects for future work.}

Before we review existing attacks on AI, we describe the existing ML threat model commonly used for these attacks.

\begin{table*}[t]
\caption{Threat models for AI security. Below, we list the attacker's knowledge, capabilities, and goals. For each attack, we denote which knowledge in terms of training data (\trainx,\trainy), test data (\testx,\testy), parameters (\params), and classifier's outputs (\classifier$(\params,\sample)$) are required. Concerning capabilities, we denote whether the attacker can alter training (\sample) or test ($\sample_t$) samples, labels of samples (\lab), or observe the output of the model (\classifier$(\params,\sample)$). For all properties, we denote required (\yes), sometimes required (\sometimes), and not required (\no). We then denote with \checkmark whether the goal of the attack is availability (Av.), integrity (Int.), or confidentiality (Conf.).}\label{tab:threatModelAcademia}
\resizebox{\linewidth}{!}{\begin{tabular}{lllllllccccl}
\toprule
& \multicolumn{3}{c}{Knowledge} & \multicolumn{4}{c}{Capabilities}& \multicolumn{4}{c}{Attacker's goal} \\
\cmidrule(lr){2-4}\cmidrule(lr){5-8}\cmidrule(lr){9-12}
& \trainx,\trainy & \testx,\testy & \params & \sample & \lab & $\sample_t$ & \classifier$(\params,\sample)$ & Av. & Int. & Conf. & Description \\ 
\midrule
Poisoning, bilevel~\cite{cina2022wild}                   &\yes             &\no         &\sometimes         &\yes             &\sometimes  &\no   &\no & \checkmark & && Decrease performance           \\
Poisoning, label flip~\cite{cina2022wild}                   &\yes             &\no         &\sometimes         &\no             &\yes  &\no   &\no & \checkmark &&& Decrease performance              \\
Backdoor~\cite{cina2022wild}                    &\yes             &\yes          &\sometimes              &\yes             &\sometimes        &\yes    &\no    && \checkmark && Misclassify samples with trigger            \\
Evasion, white-box~\cite{DBLP:journals/corr/SzegedyZSBEGF13}                &\no             &\yes          &\yes          &\no             &\no      &\yes    &\no  &&\checkmark & &  Misclassify perturbed sample     \\
Evasion, black-box~\cite{mahmood2021back}                &\no             &\yes          &\no          &\no             &\no      &\yes    &\sometimes  &&\checkmark & &    Misclassify perturbed sample     \\
Model Stealing~\cite{oliynyk2023know}            &\no         &\sometimes           &\no           &\no          &\no          &\sometimes   &\yes   &&& \checkmark & Copy model without consent                  \\
Mem. Inf.~\cite{jegorova2022survey}             &\no            &\yes            &\no          &\no           &\no        &\sometimes      &\yes       &&& \checkmark & Infer sample membership  \\
Attribute Inf.~\cite{jegorova2022survey}            &\no            &\no            &\yes          &\no            &\no        &\no      &\no     &&& \checkmark & Infer training data attributes                \\

\bottomrule

\end{tabular}}
\end{table*}

\subsection{Threat Modelling Artificial Intelligence}\label{sec:threatModelAca}
In general, we distinguish three different aspects defining an attacker's behavior, its \emph{knowledge},  \emph{capabilities}, and \emph{goal}~\cite{biggio2018wild}. We summarize these aspects of AI attacks from academic literature in Table~\ref{tab:threatModelAcademia}, and first review the properties before we discuss different attacks in the following subsection.

\textbf{Knowledge.} This aspect describes what the attacker \emph{has access to} or \emph{has knowledge about}. The training (\trainx,\trainy) or test (\testx,\testy) data are examples of information the attacker might have. In some cases, knowing this data roughly may suffice: when the victim is training an image classifier, some images can be sufficient to mount an attack, but not the same images the victim trained on are required. Independent of the data, the attacker may know the model's parameters (\params).

\textbf{Capabilities.} In contrast to knowing or observing system properties, 
threat modeling also describes what an attacker can \emph{alter}. The attacker may, for example, change input samples at training (\sample) or test time ($\sample_t$) or both. In case we are dealing with ML or RL, it might be relevant to distinguish samples (\sample) and labels (\lab), e.g., inputs and associated classes or desired behaviors. Lastly, the attacker may feed the model inputs and observe the corresponding outputs (\classifier$(\params,\sample)$). 

\textbf{Attacker's goal.} There are three principal goals~\cite{biggio2018wild}. Harming \emph{availability} decreases overall performance to a degree where this system may not be usable anymore. Targeting \emph{integrity} preserves the original performance, but specific outputs may be processed incorrectly (e.g., misclassified).  The third, \emph{confidentiality}, concerns the intellectual property of the model
and the secrecy of the training data.

\textbf{Practical concerns - 3rd parties.} In research, knowledge and capabilities are often binary (present/not present). In practice, there may be different access levels. In our questionnaire, we consider two levels of accessibility: On the one hand within a company, encompassing employees and clients; and on the other hand 3rd party, or anyone. 

Finally, cost-driven assessments should be an additional dimension in analyzing ML security in practice~\cite{apruzzese2022real}. As both attackers and defenders operate with a cost/benefit mindset~\cite{wilson2014some}, attacks will only be conducted if their benefit exceeds the costs. Similarly, defenses will only be applied if their implementation costs are lower than the respective attack’s monetary impact that materializes with a certain likelihood~\cite{ISO23894}. 

\subsection{AI Security}\label{sec:AISecThreatModelsAca}
Most AI security work focuses on ML. We thus introduce the ML-security threat models and then discuss the same attacks on other paradigms like RL and DM. We start with training time attacks like poisoning and backdoors and then discuss test time attacks like evasion, and attacks breaching confidentiality like model stealing, membership inference, and data extraction. We focus on the attacks of AI security that received the most attention and visualize them in Table~\ref{tab:threatModelAcademia}. 

\textbf{Poisoning.} In poisoning, the attacker alters training data~\cite{rubinstein2009antidote} or labels~\cite{biggio2011support} to decrease accuracy, thus targeting availability. Attacks that uniquely target labels are called label-flip attacks~\cite{biggio2011support}, whereas poisoning based on the bilevel formulation alters only samples or samples and labels~\cite{cina2022wild}.
Alternatively, in sloth attacks, the goal is to increase the model's runtime~\cite{cina2022energy}. Defending poisoning is well understood~\cite{cina2022wild}. Poisoning attacks~\cite{behzadan2017vulnerability} and defenses~\cite{RajeswaranGRL17} have been studied on RL, DM~\cite{mei2015security}, clustering~\cite{xiao2015feature}, principal component analysis~\cite{rubinstein2009antidote}, or feature selection algorithms~\cite{xiao2015feature}. 

\textbf{Backdoors.}  An alternative attack during training time are backdoors. Backdoors are chosen input patterns that reliably trigger a specified classification output, harming integrity.
There are several ways to introduce backdoors~\cite{cina2022wild}, via the training~\cite{geiping_witches_2020} or the fine-tuning data~\cite{shafahi_poison_2018}. Alternatively, a backdoored model can be provided~\cite{Doan_2021_ICCV}.
Mitigating backdoors has led to an arms race~\cite{tan2019bypassing}, where proposed defenses are broken, leading to new, stronger attacks which again have to be mitigated~\cite{cina2022wild}.
Backdoors have also been studied on RL~\cite{kiourti2020trojdrl}.

\textbf{Evasion/adversarial examples.} 
Evasion decreases the test-time accuracy of a trained and otherwise well-performing classifier~\cite{Dalvi:2004:AC:1014052.1014066,DBLP:journals/corr/SzegedyZSBEGF13}, and thus also target integrity. To this end, the attacker needs access to the test data and knowledge about the model for white-box attacks, as visualized in Table~\ref{tab:threatModelAcademia}. An exception are black-box attacks, which only require access to the model outputs and knowledge about the rough nature of the data~\cite{mahmood2021back}. Alternatively, an attack can be computed on one model $\classifier_1$ and then transferred to a second classifier $\classifier_2$ to which the attacker does not have access~\cite{papernot2017practical}. Recent works emphasize the need to correctly evaluate defenses~\cite{croce2021robustbench,tramer2020adaptive}.
Evasion has also been introduced~\cite{lin2017tactics} and tentatively defended~\cite{mandlekar2017adversarially} on RL and on clustering algorithms~\cite{joslin2019measuring}.

\textbf{Model stealing.} In model stealing, the attacker has black-box access to an ML model and copies its functionality without consent of the model's owner~\cite{DBLP:conf/uss/TramerZJRR16} and thus harms confidentiality, as visualized in Table~\ref{tab:threatModelAcademia}. Most model stealing attacks require submitting specific test queries~\cite{oliynyk2023know}, and only one paper~\cite{papernot2017practical} obtained models by labeling data from the task the model was purposely used for. In general~\cite{oliynyk2023know}, model stealing attacks are measured by the number of queries they need and how faithful they reproduce the original model. 
Similar to model stealing attacks is model extraction, where specially crafted inputs allow the attacker to deduce architectural choices like the usage of dropout~\cite{oh2019towards,jegorova2022survey}. 
Analogous to previous attacks, defenses have been proposed against both attacks, but are caught in an ongoing arms race~\cite{oliynyk2023know}. RL models can also be stolen~\cite{chen2021stealing}. 

\textbf{Membership inference.}
The following attacks target the privacy of the used training data at test time~\cite{he2022membership,jegorova2022survey}. For example, membership inference~\cite{shokri2017membership,choquette2021label} predicts membership to the training data for an existing sample based on the target model's output. To this end, attacks rely on membership metrics~\cite{shokri2017membership} or shadow-models~\cite{shokri2017membership} trained on known membership outputs. Alternatively, repeatedly querying the victim is possible~\cite{choquette2021label}. For all of these attacks, defenses have been proposed~\cite{jegorova2022survey}. Membership inference has also been demonstrated to work on RL~\cite{gomrokchi2023membership}.
Beyond membership, inversion attacks attempt to regenerate the training data based on a generative model trained with the victim's outputs~\cite{yang2019neural}. Intuitively, training the model encompasses a large amount of labeled data, which may differ from the original training data.

\textbf{Attribute inference.}
In contrast, in attribute inference, the attacker is interested in a specific sensitive attribute or feature. These attacks are mounted assuming white-box knowledge of the victim and using the weights for a meta-classifier~\cite{ateniese2015hacking}.
For these attacks, defenses have been proposed~\cite{jegorova2022survey}, but to the best of our knowledge, no works study attacks on RL or DM.

\section{Methodology}\label{sec:methodology}
Having gained an overview of the existing threat models in AI security research, we can now design a questionnaire and decide on a target group to recruit from to assess practical AI threat models. In this section, we first describe the questionnaire design and content, the pretests, and the recruiting procedure, and conclude with the sample description.

\subsection{Measuring Threat Models in Practice}
In the previous section, we discussed the attacker's knowledge, capabilities, and goals. To assess threat models in practice, we consider the threat model in Table~\ref{tab:threatModelAcademia} and determine whether knowledge and capabilities are reasonable assumptions in practice. For example 
Poisoning~\cite{cina2022wild} and backdoor attacks~\cite{cina2022wild} require access and knowledge of training data. Backdoors additionally require the submission of test data to exploit the backdoor. 
To validate these threat model requirements, we 
ask industrial practitioners whether a 3rd party can access training inputs and deployment inputs.
For backdoors, we ask additional questions about model re-use, a common assumption~\cite{shafahi_poison_2018,Doan_2021_ICCV}. 
For evasion~\cite{Dalvi:2004:AC:1014052.1014066,DBLP:journals/corr/SzegedyZSBEGF13,mahmood2021back}, knowledge and access to test data are necessary, while access to the model is optional. We thus also inquire about access to the model from our participants. This access is also implicitly relevant to an attack like model stealing~\cite{DBLP:conf/uss/TramerZJRR16,oliynyk2023know}. If access to the model is possible, the attack is superseded.
To conclude, the combinations of access control responses help us to determine whether an attack is possible (and, for model stealing, is necessary).

\subsection{Questionnaire Design}\label{subsec:questionnaire}
In other words, we focused on questions concerning the accessibility of the model and data, as these are essential components of threat models (Sect.~\ref{sec:threatModelAca}). Furthermore, previous work indicated access to models and data as a limiting factor~\cite{grosse2022so}.
As questions can be sensitive, we opted for an anonymous survey with 43 questions. 
For fast completion, 
the questionnaire only contains 
multiple choice questions, checkboxes, and relevance rankings based on a Likert scale. Questions, descriptions, and the wording of answer options for multiple-choice questions were based on prior research. In the following, we detail references used for the questionnaire along its three parts, (1) demographics, (2) AI projects, and (3) AI security. The complete questionnaire can be found in the Appendix.

\myparagraph{Demographics} We inquired the necessary data to compare to previous studies~\cite{grosse2022so} and populations~\cite{kaggle}, including gender, age, educational background, company size, AI experience, industry areas\footnote{\url{https://en.wikipedia.org/wiki/Economy_of_the_United_States_by_sector}} and team size~\cite{serban2022adapting}. We inquired about our participants' location based on dial codes to obtain privacy-preserving groups. These groups consisted of North, Central, and South America, North/Central, South, and East Europe, Africa, North, East, South/Central, and West Asia (with the Arabian Peninsula), and Australia and Oceania. 

\myparagraph{AI projects} In this part, we asked questions about threat models like as access to model components and sizes of the used data. We also inquired about other specifications such as the need for a domain expert, time constraints of the application, other specification of requirements for ML model~\cite{nahar2022collaboration,serban2022adapting}, and the possibilities to enforce constraints on the training data~\cite{serban2022adapting}. In case the participants worked with several AI-based 
projects, we asked them to here focus on one.

\myparagraph{AI security} This final part focused on the relevance of AI security, privacy, and a self-estimated likelihood of noticing an attack. We also asked whether participants had encountered an attack and what the attack consisted of, as Grosse et al.~\cite{grosse2022so}.

\subsection{Pretests and Recruiting}\label{subsec:recruiting}
After obtaining permission from our institution's ethical review board, we performed two rounds of pretests. All pretesters had AI industry experience (including ML and RL, for example) and were from the author's private networks. \new{The testers were given the questionnaire and asked to think out loud while filling it, enabling us to spot misunderstandings and unclarities.} In the first two rounds, 6 participants (one female, five male) \new{took part}. We received minor comments; all questions except three were well understood. We improved these and retested them with four fresh testers. After incorporating the minor changes their feedback agreed on, we implemented the survey in RedCap~\cite{harris2019redcap} and started to recruit. 

We advertised the study on social media channels such as Twitter and LinkedIn and initially reached out to personal contacts. We then followed previous studies approaches~\cite{grosse2022so} 
and recruited within AI Slack communities (MLOps, MLSecOps, Pyladies) or contacted potential participants via LinkedIn. 
We did not impose specific selection criteria other than currently working with AI and did not share the questionnaire if potential participants stated to work only with, for example,  ChatGPT. In other words, we targeted practitioners directly involved with AI/ML models or data engineering. While our conclusions affect AI security, the questionnaire is independent of security questions, allowing us to draw from a broader population than previous studies~\cite{bieringer2021mental,grosse2022so}. This was confirmed as we received no feedback that the questions within the questionnaire were unknown to the participants, although we did not screen for a security background. 
We opted against paying the participants to avoid money-driven participation. Still, many participants were eager to contribute 
due to their interest in the topic.
Throughout the recruiting process, we monitored the gender ratio to ensure the sample remained representative. 

We recruited for two and a half months\footnote{Recruiting period: April, 21st to July, 6th, 2023.} and allowed inputs for one more week to allow potential latecomers to participate. In total, 271 participants filled out our survey. 

\subsection{Sample Description}\label{sec:sampledescr}
A total of 271 participants filled out our questionnaire, of which $201$ replied to all questions, and $70$ submitted only part of the questionnaire. We do not exclude participants with partial replies. Instead, we report the fraction of participants not providing a reply for the question(s) discussed. 
Before we analyze the results, we describe the individual and organizational backgrounds of our participants and establish that our sample matches the larger population of AI practitioners.

\myparagraph{Individual background of participants}
Of our $271$ participants, 76\% were male, 18.1\% female, and the remainder did not reply or did not disclose their gender. Albeit the sample is largely male, the ratio is comparable to similar studies~\cite{grosse2022so} and representative of the population of AI practitioners~\cite{kaggle}. 

The distribution of participants’ age was primarily between 25 and 44, with most being between 25 and 34 (44.3\%). As before, this distribution matches similar studies~\cite{grosse2022so,kaggle}. To maintain anonymity, we asked for our participants' locations based on dial codes grouped into twelve areas. We received at least one participant from each area, our sample thus covers the entire globe. Most participants were from Southern (19.9\%) and Northern Europe (28\%) and North America (18.8\%). The fewest participants were from Central America (0.4\%), Russia/Mongolia (0.7\%), and South America (1.1\%). 7.4\% did not provide a location.
The distribution of academic degrees, with the largest group of master degrees (46.5\%) roughly mirrors previous distributions~\cite{grosse2022so,kaggle}. 
In terms of AI background, 5.2\% were trained only, with most participants (37.3\%) having 2-5 years of working experience in AI or ML. Almost as many (35.8\%) worked for more than 5 years. Intriguingly, our distribution matches more closely the US-focused distribution than the global distribution of prior work~\cite{kaggle}, possibly showing a bias of our sample towards Western countries.
In terms of team size, most of our participants worked in teams of 6-9 (27.3\%) or 3-5 (25.5\%) people, less in small teams ($<$3, 17\%) or in teams of 10-15 (12.9\%) or even larger than 15 people (14\%). This contrasts previous studies~\cite{kaggle}, which report a quarter of their population in either very small or very large teams.

\myparagraph{Organizational background of participants}
Although three quarters (77.1\%) of our participants' companies were headquartered in North America or Europe, our sample also encompassed companies from Africa (2.2\%), Latin America (0.4\%), North (0.7\%), West (3\%), South (5.5\%) and East Asia (2.5\%) and Oceania (2.2\%).  
Of these companies, roughly every tenth (9.2\%) was in automotive or a supplier of automotive, about every eighth in cybersecurity (13.7\%), and roughly every seventh in healthcare (15.5\%). Other areas encompassed education (3.3\%), arts and entertainment (3,3\%), and finance and insurance (4,8\%). The remainder were other areas.
Concerning company size, most participants were from small companies ($<$50 employees, 34\%). Second most were employed at large companies ($>$1,000 employees, 28.4\%), the remainder were in between, coherent with previous studies~\cite{grosse2022so,kaggle}.  AI maturity also coincided with previous samples~\cite{grosse2022so, kaggle}: Few (4.4\%) participants stated to work indirectly with AI, most (51.7\%) had models in production. Significantly fewer (17.7\%) were getting models into production, starting development (11.3\%), or evaluating use cases (7\%).

\section{Results}\label{sec:results}
Knowing that our sample matches the underlying population, we analyze the information our participants provided. We start with the overall threat surface, match the individual attacks, and then report findings beyond specific attacks that help to reconcile security research and practical usage of AI.

\myparagraph{Overall threat surface} We start with an overview of common access patterns to model and data as reported by our participants. The most frequent is to allow querying the model and obtaining the corresponding results, or not to give any access to either data, model, or outputs.

\myparagraph{Attack specific threat models} We then investigate the six attacks explained in the Background section. As some share significant parts of their threat models, we discuss them together. For example, poisoning and backdoors both occur during training (Sect.~\ref{sec:trtimeatt}). All other attacks take place at test time (Sect.~\ref{sec:testtime}), where membership inference and attribute inference are grouped as they are both related to privacy (Sect.~\ref{sec:privacy}). For each attack, we compare if the scientific threat model matched our sample's statistics. This was always the case, although there were also significant mismatches.

\myparagraph{AI security beyond specific attacks}. Finally, we analyze specific details affecting AI security: dataset sizes, involvement of domain experts, real-time requirements (Sect.~\ref{sec:practChallenges}), library usage (Sect.~\ref{sec:otherAttacks}), and factors related to the access to AI systems (Sect.~\ref{sec:threatmodelfactors}). As for the individual attacks, we find both alignments as well as significant mismatches.

\begin{table}
\centering
\caption{Most frequent knowledge or capabilities available within our sample. For the components training (\trainx,\trainy) and test queries (\testx,\testy), model weights (\params), and model outputs (\classifier$(\params,\sample)$), we depict access for 3rd party (\yes), and no access (\no). \new{25.1\% of replies did not cover all questions and are not included, rare combinations with <1.5\% are not listed either.}}\label{tab:threatModelInd}
\centering
\begin{tabular}{llclc}
\toprule
& \trainx,\trainy & \testx,\testy & \params & \classifier$(\params,\sample)$ \\ 
\midrule
32.8\%    &\no             &\yes         &\no         &\yes    \\
25.1\%    &\no             &\no         &\no         &\no    \\
7.4\%    &\no             &\yes         &\no         &\no    \\
2.6\%    &\yes             &\yes         &\no         &\yes    \\
1.8\%    &\no             &\yes         &\yes         &\yes    \\
1.8\%    &\yes             &\yes         &\yes         &\yes    \\
\bottomrule
\end{tabular}
\end{table}

\subsection{Overall Threat Surface}\label{sec:genthreatmodel}
Before the individual attacks, we describe the overall threat surface using frequent access patterns within our sample.

\myparagraph{Threat surface}\label{sec:accesses} We compared the threat models commonly used in research, as described in Table~\ref{tab:threatModelAcademia} in Section~\ref{sec:threatModelAca}, with the replies of our participants in Table~\ref{tab:threatModelInd}. In this table, we reported access in practice to training data (\trainx,\trainy), test data to query (\testx,\testy), the model's parameters (\params), and the classifier's outputs (\classifier$(\params,\sample))$ (\question{23},\question{24},\question{32},\question{33}). We are interested in \emph{frequent combinations} of allowed access. 68 participants, or  25.1\%, did not provide a reply in at least one field. The remainder replied to all four questions. Almost a third of the participants (32.8\%) gave access to test queries and model outputs. The second largest combination, with 25.1\%, gave no access to data, queries, models, and outputs. The third largest group with 7.4\% allowed queries only. Smaller groups contained diverse combinations, including access to everything except the model (2.6\%), all but the training data (1.8\%), or everything (1.8\%). Rare combinations included two cases where nothing but the training data was accessible, and one case with all available except outputs. We now discuss in more detail the individual attacks with their threat surface. 

\kbox{\textbf{Take away--Overall threat surface.} The most frequent access patterns are access to queries and query outputs (32.8\%) or no access at all (25.1\%).}

\subsection{Attack Specific Threat Models}\label{sec:specificAttacks}
Having described the overall practical attack surface, we now focus on the individual attacks' threat models. We study the six attacks described in Sect.~\ref{sec:AISecThreatModelsAca}, which we regroup to take into account threat model similarities. We start with training time attacks and review both poisoning and backdoor attacks. Afterwards, we focus on test-time attacks like evasion and model stealing and finally discuss privacy breaching attacks like membership and attribute inference. 

\subsubsection{Training-Time Attacks}\label{sec:trtimeatt}
Both poisoning and backdoor attacks perturb the training data to affect the resulting model (compare Table~\ref{tab:threatModelAcademia}). Consequently, the question is how often the training data is accessible (\question{23}). Of our participants, 71.6\% reported that the training data was not accessible, and 6.6\% that the data was publicly accessible.

These numbers reflect access to the final training data---it might still be possible to tamper with the data at its public origin; when data comes for example from the internet. To this end, we investigated combinations of inaccessible training data (\question{23}) and the percentage of training data from public sources (\question{28}).
Here, 100\% corresponds to the subset of all participants who reported that their training data was not accessible.
The largest group (47.1\%) kept their data inaccessible and did not use any data from public sources. Yet, 6.6\% stated that 1\%-5\% of their training data came from public sources. The same held for 5\%-10\% (9.1\%), 10\%-15\% (4.1\%), and 25\%-50\% (5\%) training data from public sources \new{(of our participants)}. Also, higher percentages like 50\%-75\% (7.4\%) or higher than 75\% (10\%) of the training data were from public sources even if the resulting data was inaccessible, outlining the need for a complex consideration of practical data security risks.

On the other hand, only 18\% of our participants reported that more than 50\% or an unknown amount of the data stemmed from public sources. 
This may indicate that from a practical point of view, relying on high percentages of clean data for defenses is possible. Yet, data quality may then be a problem, and this may be a poor security design choice.

\begin{table}[t]
\caption{Comparing assumptions about alterable training data in poisoning~\cite{cina2022wild} and backdoors~\cite{cina2022wild} to our sample. We state the percent of training data that can be altered, the amount of poisoning and backdoor papers with this specific assumption. Finally, we show the percentage of participants in our sample with this amount of alterable data. The percentages marked with $^*$ were misaligned with our questionnaire and were thus estimated. \new{There were 20.7\% missing replies in this question.}}\label{tab:poisoning}
\centering
\resizebox{\linewidth}{!}{\begin{tabular}{lllrr}
\toprule
Percent training & \# Poisoning  & \# Backdoor & Our \\ 
data altered & papers~\cite{cina2022wild,carlini2021poisoning} & papers~\cite{cina2022wild} & findings \\ 
\midrule
$>$30\%   & 3                        & ---                           & $^*<$30.3\%                     \\
10-30\%              & 13                        & 20                        & $^*<$10.7\%                      \\
$<$10\%      & 1                            & 12                        & 20.4\% \\
$\emptyset$   & ---  & --- & 30.3\% \\
\bottomrule
\end{tabular}}
\end{table}

\myparagraph{Poisoning}\label{sec:poisoning} Cinà et al.~\cite{cina2022wild} surveyed the percentage of training data an attacker altered in poisoning attacks. Albeit their analysis focused on vision tasks, their overview summarized common attack assumptions in poisoning. Of their 16 analyzed poisoning papers, the majority (13) tampered with 10-30\% training data. The remaining 3 papers altered even more data. We compared these numbers to the percentage of training data from public sources (\question{28}), knowing the amount of accessible training data was low. In our sample, either a large fraction of the data came from public sources or none. This contrasts with the existing poisoning threat models, where most papers studied a setting altering 10-30\% training data. Many poisoning papers thus studied settings that were rare in practice according to our sample.

\begin{figure}[t]
 \centering
\includegraphics[width=0.475\textwidth]{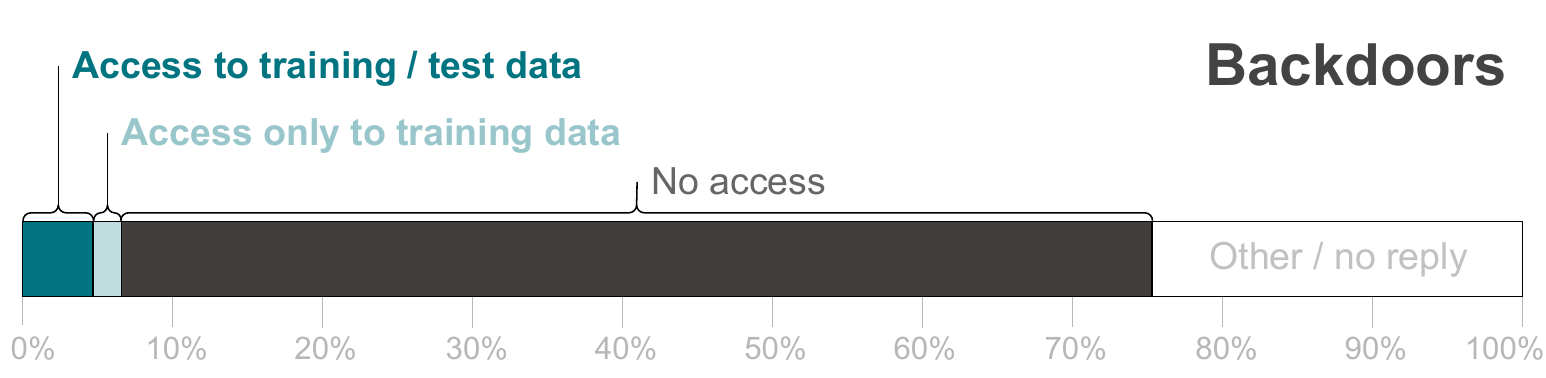}
\caption{Backdoor threat model in percent of our participants' replies. We report 3rd party access: White denotes incomplete data or an irrelevant threat model (e.g., only test data accessible). Black represents no access, \color{mycolor}{turquoise}\color{black}{ the backdoor threat model.}\color{light-my}{ Light turquoise}\color{black}{ denotes insufficient access for backdoors, but sufficient access for poisoning attacks.}}\label{fig:backdoor}
\end{figure}

\myparagraph{Backdoors}\label{sec:backdoor} Analogous to poisoning attacks, the same survey~\cite{cina2022wild} also covered backdoor attacks. Of 32 systematized papers,
about two-thirds (20) tampered with 10-30\% of the training data. While no paper altered more data, the remaining 12 papers perturbed less than 10\% data. 
As before, we compared these results to the percentage of training data from public sources (\question{28}). As before, the heavily studied middle range (10-30\%) was the least common in practice.

To exploit the backdoor, the attacker must access the test data. We thus investigated combinations of training and test data access within our sample and visualized the results in Figure~\ref{fig:backdoor}. Of our participants, 6.6\% reported training data was accessible to a 3rd party. However, adding the constraint of accessible test data, this reduced to 4.7\%; a low attack surface towards backdoors. The setting where only training data is available is studied in poisoning or triggerless backdoors, which instead target a small group of clean samples~\cite{cina2022wild,geiping_witches_2020}. Of 32 papers, 10 rely on this specific threat model~\cite{cina2022wild}.

Another assumption in backdoor attacks is that practitioners rely on existing models and fine-tune these. We combined the information provided by Cinà et al.~\cite{cina2022wild} about the fine-tuning setting and our participants' replies (\question{21}). Of the 32 backdoor papers, 12 dealt with a fine-tuning setting, e.g., the victim took an external model and fine-tuned this model on internal data. Almost half of our participants (48.1\%) stated to use third-party models and then fine-tune them. Only about a quarter denied using any third-party models (24.3\%). This setting was studied in 12 (37.5\%) of the backdoor papers. These findings highlight the need to study security risks both for pre-trained and end-to-end training, as is currently the case. Furthermore, backdooring or poisoning a model used later on circumvents the need to alter training data.

\myparagraph{Discussion} While there are notable exceptions of papers assuming very small poisoning/backdoor percentages of less than 3\% in vision~\cite{carlini2021poisoning,han2022physical}, object detection~\cite{ma2022dangerous}, and point clouds~\cite{xiang2021backdoor}, more such work is needed. Furthermore, there are two limitations to discuss. On the one hand, we currently do not know which quality checks are put upon public data, and how this affects current attacks. \new{In addition, an attack altering 5\% of the training data may affect 20.4\% of our participant's models, as the true allowed percentage may be lower. However, the attack affects 40.6\% of the cases, since \emph{at least} 5\% data access is required.}

\kbox{\textbf{Take away--Training time attacks.} We find evidence that assumptions of poisoning and backdoor threat models are met in practice. Yet, while data can often not be accessed directly, poisoning and backdooring may be executed via public data sources. Our participants also reported frequent (about 50\%) use of third-party models which are then fine-tuned.}

\begin{table}[t]
\caption{Comparing assumptions about  queries in black-box evasion~\cite{mahmood2021back} and model stealing~\cite{oliynyk2023know} papers. We state the number of queries that can be submitted, then the amount of black-box evasion and model stealing papers assuming the specific amount. Finally, we state our participants reported query amounts, \new{with 19.1\% missing replies.} }\label{tab:testQueries}
\centering
\resizebox{\linewidth}{!}{\begin{tabular}{lllrr}
\toprule
Possible   & \# Evasion black-    & \# Model st. & Our   \\ 
queries   & box papers~\cite{mahmood2021back}    & papers~\cite{oliynyk2023know} & findings    \\ 

\midrule
$\emptyset$              & $^*$ & ---          & 36.5\%  \\
$<$100     & 2              & 5           & 15.5\% \\
100-1k             & 8              & 9           & 4.8\%                                                                    \\
1k-100k            & 1              & 16          & 7.4\%                                                                    \\
$>$100k & ---            & 10          & 1.1\% \\
$\infty$  & 11  & 40 & 15.6\% \\
\bottomrule
\end{tabular}}
\end{table}

\subsubsection{Test-Time Attacks}\label{sec:testtime}
Evasion, model stealing, and privacy-based attacks target a model at test time (as visualized in Table~\ref{tab:threatModelAcademia}). They are thus similar as they require the submission of test inputs and observing the model's outputs. Before we cover these attacks individually, we examine these requirements in general.

In terms of test data access (\question{33}), almost half (48.1\%) of our participants reported that the model could not be queried. On the other hand, 39.5\% reported that querying their model was possible. The model itself (\question{24}) was not accessible for three-quarters (75.5\%) of our participants, \new{for} 7.7\% \new{of them}, the model was publicly available. Model outputs (\question{32}) were available more readily: outputs were not accessible in roughly a third (37\%), and freely available in half (49.1\%) of the cases.

To understand how many queries could be submitted at test time (\question{33}), we briefly report statistics. In most (36.5\%) cases, no queries were possible. This is followed by less than 100 queries (15.6\%) and infinitely many (15.5.\%), followed by 1,000 - 100,000 (7.4\%). The least frequent are more than 100,000 queries (1.1\%) and 100-1,000 queries. 
  
\kbox{\textbf{Take away--Test time attacks.} Compared to the training data, the threat surface is larger at test-time but queries to accessible models are either very constrained or unconstrained.}

To be able to cover each attack's specialties, we now analyze the specific threat models individually.

\begin{figure}[t]
 \centering
\includegraphics[width=0.475\textwidth]{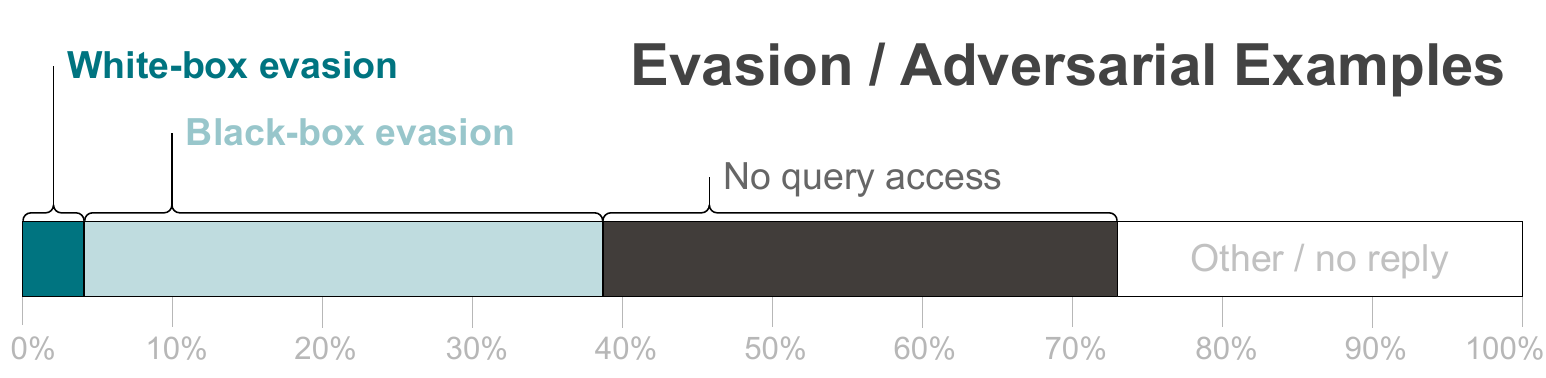}
\caption{Evasion threat models in percent of our participants' replies. We report 3rd party access: White denotes incomplete data or an irrelevant threat model (e.g., only model accessible). Black represents no access, \color{mycolor}{turquoise}\color{black}{ white-box and }\color{light-my}{light turquoise}\color{black}{ black-box evasion} threat models.}\label{fig:evasion}
\end{figure}

\myparagraph{Evasion}\label{sec:evasion}
Many evasion attacks assume access to the model and the model's inputs at test-time to alter predictions~\cite{biggio2018wild,Dalvi:2004:AC:1014052.1014066,gnanasambandam2021optical,madry2018towards} (see also Table~\ref{tab:threatModelAcademia}). We examined these threat models and visualized our participants' replies in Figure~\ref{fig:evasion}. 
We found that 3rd party access to these two features (\question{24} and \question{33}) was rare and only reported by 4.1\% of our participants. If we dropped the white-box constraint and permitted the attacker to have no access to the model, this percentage increased strongly to 34.6\%. As expected, black-box attacks could be carried out more frequently in our sample.

We thus focus on black-box attacks~\cite{biggio2018wild,mahmood2021back,croce2021robustbench,garcia2023less} and the number of queries needed for an attack (\question{33}). For the sake of this comparison, we relied on the overview of Mahmood et al.~\cite{mahmood2021back} and ignored whether attacks are targeted or untargeted and whether hard or soft labels are required. We report the minimal empirical amount of queries documented by Mahmood et al.~\cite{mahmood2021back} in Table~\ref{tab:testQueries}. Few (2) papers operated in the setting \new{most frequently reported} (15.6\%) with less than 100 queries allowed. Most papers (8) needed  100-1,000 queries, which is the range least often (4.8\%) reported by our participants. One paper required 1,000-100,000 queries, which is slightly more frequent (7.4\%). On the other hand, 15.5\% of our participants stated to allow infinitely many queries. In this sense, access to AI systems in practice was all-or-nothing, with few test queries or infinitely many. Research, in contrast, focused on the middle amount of queries, possibly as a consequence of decreasing the number of queries needed. An in-depth understanding of the required queries to attack a model is subject to ongoing research~\cite{garcia2023less}. In addition, our work is a call for transferability studies, when neither model nor data are known, as uttered by Sheatsly et al.~\cite{sheatsley2022space}. Such a setting (attacking only via test data) was most practical according to our participants.

\kbox{\textbf{Take away--Evasion.} According to our participants, 4.1\% of their models were  vulnerable against white-box evasion.
Often, the model is not available; and either very few or an unconstrained number of queries is granted, whereas research assumes a moderate query number. Finally, in some cases only data can be submitted without model feedback, 
 highlighting the need to deepen our understanding of transferability. }

\begin{figure}[t]
 \centering
\includegraphics[width=0.475\textwidth]{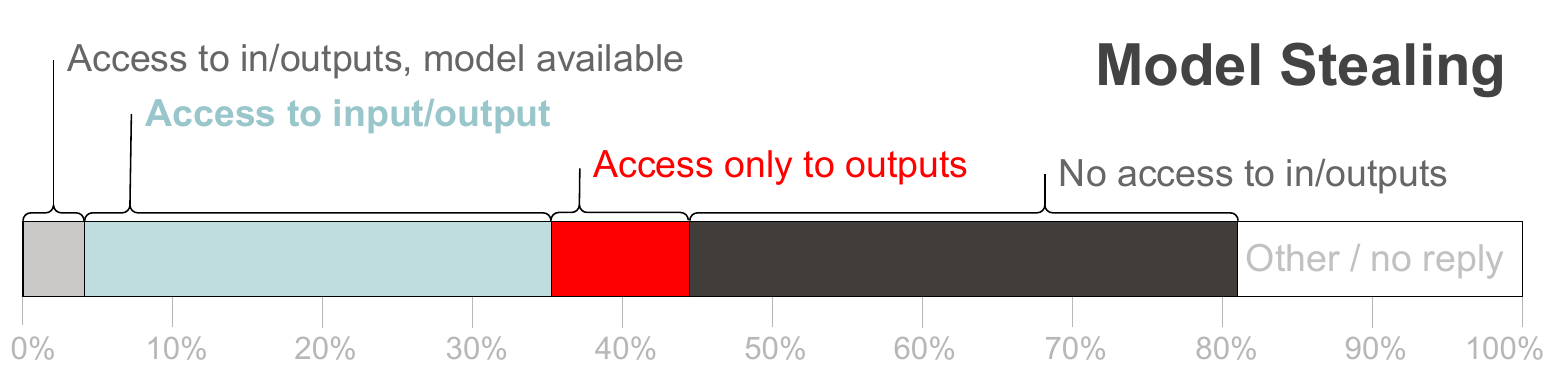}
\caption{Model stealing threat model in percent of our participants' replies. We describe 3rd party access: White denotes incomplete data or an irrelevant threat model (e.g., only test inputs are accessible). Black represents no access, \color{mycolor}{turquoise}\color{black}{ denotes the academic threat model, }\color{mygray}{gray}\color{black}{ that the attack is obsolete as the model is available.} \color{myred}{Red}\color{black}{ denotes a rarely studied threat model in current research.}}\label{fig:modelstealing}
\end{figure}    

\myparagraph{Model stealing}\label{sec:modelstealing}
Model stealing attacks target the model via test inputs and outputs~\cite{oliynyk2023know}  (\question{24}, \question{32}, and \question{33}). The goal is to obtain a copy of the target model in terms of functionality or a direct copy of the weights. We examined this threat model and plotted the corresponding percentages in Figure~\ref{fig:modelstealing}. 
44.5\% of our participants reported that they allowed public access to model outputs. Most model stealing attacks~\cite{oliynyk2023know,DBLP:conf/uss/TramerZJRR16} require submitting specific queries, decreasing this percentage to 35.3\%. 
In 4.1\% \new{of our sample}, the model itself was however also accessible, defeating the purpose of the attack. 
Although the assumptions of model stealing are met in some cases, in about 10\% of the cases, it would be beneficial to study model stealing attacks that are purely based on observing the outputs of samples that are not under the attacker's control, as somewhat studied by Papernot et al.~\cite{papernot2017practical}.

An additional factor in model stealing is, as before, the submittable number of queries to the target model (\question{33}). We compared the number of queries reported by Oliynyk et al.~\cite{oliynyk2023know} to our sample in Table~\ref{tab:testQueries}. Most of the 40 papers surveyed required between 100 and 100,000 queries, the numbers our participants reported the least frequently. Only five papers relied on less than 100 queries and aligned with a larger (15.6\%) percentage within our sample. 
We further investigate the relationship between the number of queries allowed and model complexity (as approximated by input size, \question{29}). There is no statistically significant correlation.  
The most frequent combinations of replies were with 10.7\% inputs of size 100-1k, 9.3\% 10-100, and 6.3\% no applicable feature size, each with less than 10 queries. 
Both an input size of 10-100 with 10-100 queries and not applicable input size with unconstrained inputs \new{were reported in our sample} in 4.1\%. All other combinations appeared ten times or less in the responses, with 24.4\% responses not being analyzed due to missing data.

\myparagraph{Discussion} \new{As before, an attack needing $<$100 queries possibly applies to 15.6\% of the models within our sample, but also in 28.8\% (4.8\%+7.4\%+1.1\%+15.6\%, see Table~\ref{tab:testQueries}) of the cases, as \emph{at least} 100 queries are required.}

\kbox{\textbf{Take away---Model stealing.} Model stealing can be carried out in practice. Yet, in some cases where input and output are accessible, the model is accessible, too. According to our sample, a relevant setting for model stealing attacks is only output visibility, without the possibility of submitting test queries. In addition, most attacks study infrequent numbers of queries, as either more or fewer samples are granted commonly. More work is needed to understand the relationship of amount of queries and model complexity in practice.}

\begin{figure}[t]
 \centering
\includegraphics[width=0.475\textwidth]{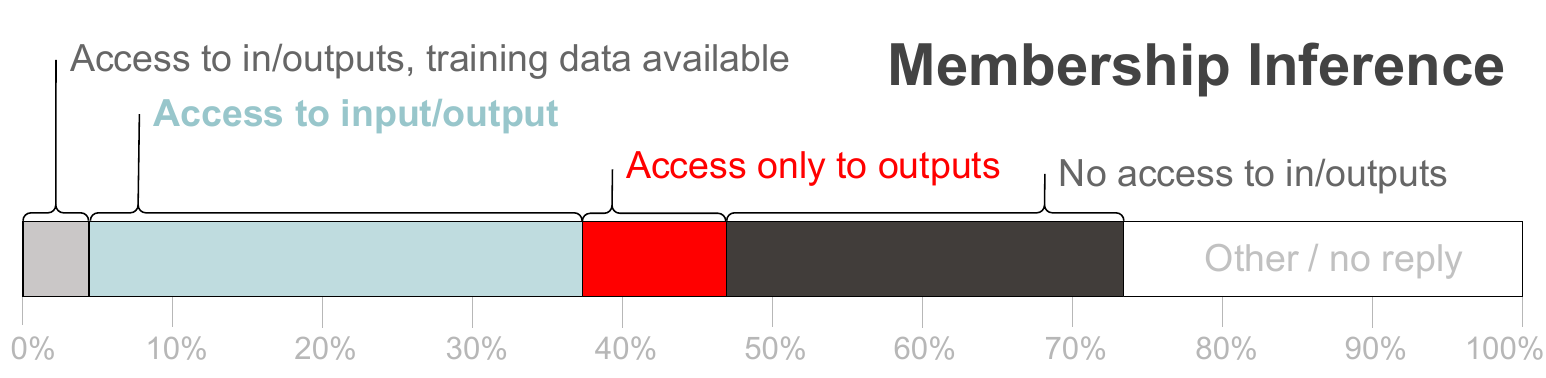}
\includegraphics[width=0.475\textwidth]{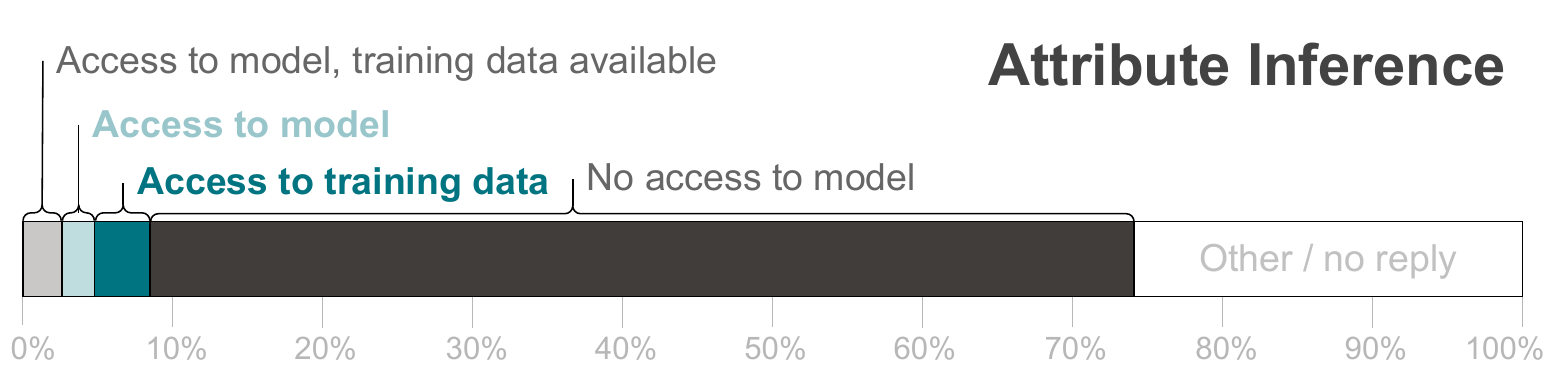}
\caption{Membership and attribute inference threat models in percent of participants' replies. We describe 3rd party access: White denotes incomplete data or irrelevant threat models, black represents no access, \color{light-my}{ turquoise}\color{black}{ denotes existing threat models, }\color{mygray}{gray}\color{black}{ means that the attack} is obsolete as the training data is available, too. For membership, \color{myred}{red}\color{black}{ denotes a threat model not studied so far.} In the case of attribute inference, \color{mycolor}{turquoise}\color{black}{ denotes no model access, but the property can directly be inferred from the training data.}}\label{fig:privacy}
\end{figure}

\subsubsection{Privacy Attacks}\label{sec:privacy}
We here discuss the two attacks inferring training data properties, first membership inference and then attribute inference. 

\myparagraph{Membership inference}\label{sec:membership}
Membership attacks use test queries and their corresponding output from the target model to infer information about the training data~\cite{he2022membership,jegorova2022survey} (\question{23},\question{32},\question{33}; see Table~\ref{tab:threatModelAcademia}). We visualize the practical threat models at the top of Figure~\ref{fig:privacy}.
About half of our participants allowed 3rd party access to their model outputs. When combined with accessible test data, this decreased to 37.3\%. In 4.4\% of these cases, the training is then public, too. 
Independently, most membership attacks~\cite{he2022membership} assumed one input and output per training point to determine membership. As the number of queries was often less than 100 with only outputs visible, it would be beneficial to understand if membership can be inferred for several points at once.

\myparagraph{Property inference}\label{sec:propInference}
Property inference attacks derive from the model's weights properties of the training data~\cite{jegorova2022survey} (\question{25},\question{23}; see Table~\ref{tab:threatModelAcademia}). Analogous to previous observations, 
65.7\% of the participants give no access and 4.8\% grant 3rd party access to their weights. In 2.5\% of these cases, however, the training data is then publicly available. 

\kbox{\textbf{Take away---Privacy attacks.} Membership and property inference can be carried out in practice. In some cases where threat models apply, the data was however accessible, too. According to our sample, it would be beneficial to study membership attacks with fewer attacker capabilities: only access to outputs.}

\subsection{AI Security Beyond Specific Attacks}\label{sec:beyondIndAttacks}
There is more to learn from our survey respondents than attack-specific threat models. This section presents information that either supports existing work or can be used to support future,  realistic AI threat modeling. To this end, we first discuss factors like common dataset sizes, involvement of domain experts, real-time requirements,  library usage, and finally which factors influenced given access to AI systems. 

\subsubsection{Practical Challenges for AI Security Research}\label{sec:practChallenges}
To better align AI security research and practice, we discuss relevant information to make future work in AI security more practical. To this end, we review which data types are commonly used in the industry, and then discuss practical challenges on both the attack and defense sides.

\begin{table}[t]
\caption{Summary of our participant's reported data. We denote the sample size in the number of features \sample and the size of the training set $|\trainx,\trainy|$. We also give examples of academic datasets of similar dimensions. \new{21.8\% partial replies are not listed, neither are 
combinations with a prevalence <5\%.}}\label{tab:datasizes}
\centering
\resizebox{0.95\linewidth}{!}{\begin{tabular}{llll}
\toprule
& size of \sample & $|\trainx,\trainy|$ & Example dataset \\ 
\midrule
9.6\% & 10-100 & 10$^5$-10$^8$  \\
9.6\% & 10$^2$-10$^3$ & 10$^3$-10$^5$ & (Fashion) MNIST~\cite{lecun1998gradient,xiao2017fashion} \\
7.8\% & 10-100 & 10$^3$-10$^5$ & Iris~\cite{asuncion2007uci}, Wine~\cite{asuncion2007uci}, Spam~\cite{asuncion2007uci}\\
7\%  &  10$^2$-10$^3$ & 10$^5$-10$^8$ & \\
5.5\% & 10$^3$-10$^5$ & 10$^2$-10$^3$ & CIFAR~\cite{krizhevsky2009learning}, 
Drebin~\cite{arp2014drebin} \\
5.5\% & other & 10$^3$-10$^5$ & \\
5.2\% & other & other & Open AI Gym~\cite{brockman2016openai}\\
\bottomrule
\end{tabular}}
\end{table}

\myparagraph{Common dataset sizes}\label{sec:datasets} We investigated the most frequent dataset properties in terms of feature size (\question{29}) and training set size (\question{30}) in Table~\ref{tab:datasizes}, as data dimensionality (e.g., number of features) and samples may influence security~\cite{wong2018scaling}. Considering both questions in combination, 59 participants (21.8\%) did not reply to one of the questions. Overall, the data size was very diverse. Yet, small input sizes (10-100 or 10$^2$-10$^3$) were prevalent. Only considering \question{29} about feature sizes, over fifty percent of our participants reported a small number of features (10-100 (27.2\%) or 10$^2$-10$^3$ (25.8\%)). Also, non-quantifiable data was frequent (16\%), as shown in Table~\ref{tab:datasizes}. 
These sizes match datasets such as CIFAR~\cite{krizhevsky2009learning}, CelebA\cite{liu2015faceattributes}, and Open AI Gym~\cite{brockman2016openai}. We also found dimensions of datasets that were used heavily before, including MNIST~\cite{lecun1998gradient}, Drebin~\cite{arp2014drebin}, and smaller datasets like Iris~\cite{asuncion2007uci}, Wine~\cite{asuncion2007uci}, and Spam~\cite{asuncion2007uci}. Overall, frequent datasets in practice were smaller than current academic datasets: there were no image-net~\cite{deng2009imagenet} like datasets frequent in our sample. Many practitioners worked thus with smaller data in terms of features (with a potentially large number of samples) which has, to the best of our knowledge, not been studied in depth yet.

\myparagraph{Attack and mitigation challenges}\label{sec:genattacksdefenses}
We investigate two more properties that potentially affect AI security. First, we asked our participants whether they relied on a domain expert (\question{15}). The presence of an expert may imply that also to attack, specific knowledge is required to constrain for example feature changes. A large fraction (37.8\%) of our participants reported relying on domain experts. An additional 4\% wanted to work with one but did not find someone yet. Domain knowledge should thus be considered when studying AI security.

Furthermore, we inquired whether our participants required real-time responses to their AI-based systems (\question{17}). Such a requirement affects the overall time that is available to defenses. Only 11\% of all participants reported that their applications were \emph{not} time-critical at all, but only about a third (35.8\%) required real-time results. This shows the need for mitigations to cope with time constraints practice.

\kbox{\textbf{Take away--General threat modelling.} Current research datasets match practical settings. Yet, some are in practice smaller in features than current academic counterparts, outlining the need to also study data security for a few features and many samples. Furthermore, our results emphasize the need to study constraints in terms of expert knowledge or time.}

\subsubsection{Code Libraries as a Security Factor}\label{sec:otherAttacks}
While libraries are acknowledged as a relevant security factor in other areas than AI security~\cite{prana2021out}, few works study AI security about libraries~\cite{grosse2019adversarial,shumailov2021manipulating,shumailov2020sponge, cina2022energy,hong2020panda}. We briefly state our results here as to whether our participants used libraries.

\myparagraph{Vulnerabilities via AI libraries} A recent strain of attacks relies on manipulated libraries~\cite{grosse2019adversarial,shumailov2021manipulating} to affect the order of data~\cite{shumailov2021manipulating} or the initial weights of a target model~\cite{grosse2019adversarial}. To assess the feasibility of such attacks, we inquired how models are developed in terms of code (\question{19}). While almost all participants (90.5\%) developed models using self-written code, they further relied on additional tools. Such tools included open-source code (88.8\%) or proprietary solutions (36.5\%). While the aforementioned attacks are complex as a tempered library has to be placed on the victim's machine first, it might be worthwhile to investigate whether hashcodes and other security measures are in place to prevent such attacks.

\myparagraph{Energy saving libraries}\label{sec:sloth} Several recent works increase the run-time of deep learning models under the assumption that these models use energy-saving soft- or hardware~\cite{cina2022energy,hong2020panda}. To verify that this is the case also in practice (\question{22}), we inquired about the use of energy-saving libraries, software, or self-written code. More than a third of our participants confirmed using such methods (34.3\%), with almost an additional quarter (24\%) stating that they sometimes relied on these techniques. Roughly a third (31.4\%) reported not to rely on energy saving. With the constraint that to the best of our knowledge, there is currently no understanding of whether attacks transfer between different energy-saving methods, it seems worthwhile to investigate the security of such libraries further.

\kbox{\textbf{Take away.} Libraries can be security relevant for AI.}

\subsubsection{Factors Determining Practical Threat Models}\label{sec:threatmodelfactors}
Finally, we review potential factors that influence security factors such as access control on training data and model (\question{23}-\question{24}), public fractions of training (\question{28}) and test data (\question{27}), submittable queries (\question{32}), visible outputs (\question{33}) and usage of pre-trained models (\question{21}) and reliance on domain experts (\question{15}). We tested candidate variables like AI knowledge (\question{5}), security knowledge (\question{6}), AI security knowledge (\question{7}), company size (\question{10}), AI maturity (\question{14}), team size (\question{11}) and presence of a domain expert (\question{15}) and computed the Spearman correlation to determine relationships. We set as $p$ value $0.05$ with Bonferroni correction for repeated testing, yielding a significance level of $0.05/56=0.0009$.
Most of these combinations were not statistically significant, with a few notable exceptions, as visible in Table~\ref{tab:Corrs}.

\myparagraph{AI maturity and company size}
In our sample, AI maturity affected how much test and training data came from public sources. 
For both training (-0.24, $p=4.8e^{-07}$) and test (-0.34, $p=0.0005$) data, the correlation was negative. A negative correlation indicated that more mature companies tended to collect less data from public sources. 

\myparagraph{AI security and AI knowledge} 
AI knowledge affects, analogous to AI maturity, how much training (-0.23, $p=0.00077$) and test data (-0.23, $0.0008$) were sourced from public places. As with AI maturity, this correlation was negative. This indicated that as practitioners were more knowledgeable, less data stemmed in either case from public sources.

\myparagraph{Possible influences}
A few factors in our survey affected security-relevant features like access to the model, model components, or data. A possible explanation is that these are influenced by factors not considered in this survey, for example, business models, the application, or industry area. In contrast, data collection practices are correlated to AI knowledge and the AI maturity of the company.

\kbox{\textbf{Take away.} AI security-relevant factors are not correlated to security or AI security knowledge in our sample. Both AI maturity and AI knowledge influence negatively whether data comes from public sources.}

\begin{table}[t]
\caption{Analyzing factors that influence threat model properties. We denote a negative correlation with $n$ and the absence of a statistically significant relationship with --.}\label{tab:Corrs}
\centering
\begin{tabular}{llllllll}
\toprule
& \rotatebox{90}{AI Know.} & \rotatebox{90}{Sec. Know.} & \rotatebox{90}{AI Sec. Know.}& \rotatebox{90}{Comp. Size}&\rotatebox{90}{AI Maturity} & \rotatebox{90}{Team Size}&\rotatebox{90}{Domain Expert} \\
\midrule

Access (\trainx,\trainy)  & --           & --       & --          & --      & --     & --      & --     \\
Access \params   & --           & --       & --          & --      & --     & --      & --     \\
\# queries & --       & --       & --          & -- & --   & --      & --    \\
Access \classifier$(\params,\sample)$ & --       & --       & --          & -- & --   & --      & --    \\
(\testx,\testy) from public & $n$           & --       & --          & --      & $n$  & --      & --    \\
(\trainx,\trainy) from public & $n$        & --       & --          & --      & $n $  & --      & --    \\

Pretrained \classifier & --           & --       & --          & --      & --  & --      & --        \\
Domain expert & --           & --       & --          & --      &      -- & -- &   \\
\bottomrule
\end{tabular}
\end{table}

\section{Limitations}\label{sec:limitations}
In this section, we discuss the limitations of our study. We first describe sample limitations, then proceed to discuss limitations within our questionnaire, and conclude the section by discussing methodological limitations.

\myparagraph{Sample limitations} 
Our sample is biased towards the global north, especially Europe, and is limited to English-speaking practitioners. Albeit we managed to recruit over 250 participants, we could not find reliable and consistent scientific references to estimate the global target population of industrial practitioners working with AI. However, for a population larger than 50,000, and a confidence interval of 95\%, our sample's margin of error lies around 6\%. Reducing this margin significantly to a few percent, 
 for example, 2\%, would require several thousand participants. Furthermore, in terms of demographics, our sample matches the overall population~\cite{kaggle} rather well (Sect.~\ref{sec:sampledescr}). Given that our goal is to identify conceptual mismatches of threat models in the wild compared to research, we find this margin of error acceptable. 

 \myparagraph{Questionnaire limitations} Despite our best efforts and many pretests, some questions could not be used for our analysis. This included information on the detectability of attacks (\question{40}) and questions where we inquired information about the secrecy of the input encoding (\question{18}), the expected performance (\question{34}), and the ease to assess the quality of training data (\question{35}). These questions would have helped to determine the difficulty an attacker faces when targeting a model. We relied on responses on a scale from 1 to 100. Still, the distribution of replies matched a normal distribution with a mean of 50 and quartiles around 25 and 75, indicating that the questions did not contain enough information for analysis. We leave a detailed study of these aspects for future work. \new{Orthogonally, we had planned to ask for security-relevant output transformations like not providing confidence scores. However, our pre-tests outlined that this was too specific for a sample covering RL or DM with non-discrete outputs. We thus left a study of these aspects for future work.}

\myparagraph{Methodological limitations} We did not review the entire body of AI security work. Given that there are several thousand research articles about AI security\footnote{\url{https://nicholas.carlini.com/writing/2019/all-adversarial-example-papers.html}}, this endeavor is beyond a single paper. We instead rely on surveys~\cite{cina2022energy,mahmood2021back,oliynyk2023know,jegorova2022survey,he2022membership} representing the state of the art for different attacks. We chose these surveys explicitly as they reviewed properties related to the threat models of the analyzed attacks. Some of these surveys focus on specific areas like computer vision~\cite{cina2022wild}. The scope of our comparison is thus limited and may be biased. Yet, we reason that this overview is sufficient to identify conceptual gaps. In addition, some attacks depended on factors like memorization or overfitting (membership inference)~\cite{yeom2018privacy}. As these are complex phenomena, we opted against analyzing them. While this limits our insights on these attacks, we leave this aspect for future work. Orthogonally, it is important to recognize that our study relies on self-reported properties and sheds light on what attacks are possible through channels known to practitioners. Real-world practical attack threat levels may be higher.

Independently, the practical threat models we discuss represent a momentary picture of how AI is applied in practice. Usage may change over time, resulting in evolving threat models, which should be monitored over time. Finally, AI usage is strongly dependent on the context of a specific application, which we do not cover but leave for future work.

\section{Implications and Future Work}\label{sec:implications}
Having discussed the limitations, we are ready to discuss the implications and implied future work of our study. As the most important implication of our work is directing future research in AI security, we first discuss these research directions. Afterwards, we discuss additional implications, concerning AI regulation and AI security in practice. Where applicable, we also delve into future work for these latter implications.  

\begin{table*}[t]
\caption{\new{\textbf{Main results.} All attacks exist in practice ($\exists$?) and require mitigations. We also list the threat models that practice-oriented research should focus on and denote the prevalence in our sample, where a higher prevalence may indicate a practically more relevant attack. We then summarize additional directions for research.}}\label{tab:threatModelResults}
\resizebox{\linewidth}{!}{\begin{tabular}{lrrrrr}
\toprule
 &     & \new{Relevant Practical setting}  &  Prevalence \\
Attack & $\exists$? & \new{to be researched} &  in sample & \new{Possible further research}\\
\midrule
Poisoning~\cite{cina2022wild}   & \checkmark  & \textless{}10\% training data alterable      & rare & defense trade-offs \\
Backdoor~\cite{cina2022wild}             & \checkmark  &     \textless{}10\% training data alterable      & rare & model re-use\\
Evasion, white-box~\cite{DBLP:journals/corr/SzegedyZSBEGF13}             & \checkmark    &  &    rare & (model) transferability \\
Evasion, black-box~\cite{mahmood2021back}              & \checkmark       & \textless{}100 queries possible &  $>33\%$ & (model) transferability \\
Model Stealing~\cite{oliynyk2023know}        & \checkmark  &           \textless{}100 queries                                                           &           $>33\%$  &  attacks without query access       \\
Mem. Inf.~\cite{jegorova2022survey}  & \checkmark   &             \textless{}100 queries    &   $>33\%$   & attacks without query access             \\
Attribute Inf.~\cite{jegorova2022survey}  & \checkmark && rare & attacks without model access \\                                                                                        
\bottomrule
\end{tabular}}
\end{table*}

\subsection{Future Work in AI Security}
We found several gaps between the researched threat models and practical AI usage (Sect.~\ref{sec:results}). Consequently, most of our implications translate to direct recommendations of previously overlooked aspects. In this section, we give the big picture by combining our findings for each attack, listing open questions alongside. An overview of these results can be found in Table~\ref{tab:threatModelResults}. At the end of the section, we summarize insights that go beyond individual attacks. 

\myparagraph{Poisoning and backdoors} Poisoning and backdoor threat models apply in practice (Sect.~\ref{sec:trtimeatt}). Further studies should focus on ending the arms-race and deepening our knowledge of defense trade-offs~\cite{cina2022wild}. At the same time, current percentages of frequently altered training data are not well aligned with the percentages reported by our practitioners (Sect.~\ref{sec:trtimeatt}). Although some practitioners currently report high training amounts from public sources, this is deemed to decrease as attacks or data quality problems occur. Finally, given that practitioners rely on fine-tuned pre-trained AI models (Sect.~\ref{sec:backdoor}), corresponding risks need to be assessed~\cite{hong2022handcrafted}. 

\myparagraph{Evasion} We found evidence of the applicability of (black-box) evasion threat models (Sect.~\ref{sec:evasion}), and recommend further study to end the arms-race~\cite{croce2021robustbench,tramer2020adaptive}. Still, more emphasis should be put on studying attacks that succeed without knowledge of the exact data and model outputs (Sect.~\ref{sec:testtime}). This is aligned with previous observations that ``attackers don't compute gradients''~\cite{apruzzese2022real}, as for gradients both input and output pairs are required. A similar perspective on this requirement is that more work is required on transferability. More precisely, and as stated by Sheatsly et al.~\cite{sheatsley2022space}, more work should study transferability across different datasets, not only across models. If queries are allowed, the number of queries should be minimized, ideally to less than 100, to reflect frequent settings within our sample (Sect.~\ref{sec:evasion}). Frequently, there were also no limitations on the number of queries.  Yet, adding such constraints is straightforward, and security assessments should not rely on changeable configurations.

\myparagraph{Model stealing} We found evidence of the applicability of model-stealing threat models (Sect.~\ref{sec:testtime}). Future work should address the corresponding arms-race~\cite{oliynyk2023know}. 
We further found a mismatch of used queries in model stealing and a mismatch for the attacker's capabilities overall (Sect.~\ref{sec:testtime}). Consequently, we recommend reducing used queries, and not relying on currently reported high amounts of queries, similar to evasion. In some cases,  only outputs are observable in our sample. It may thus be beneficial to understand the limitations of retrieving information only by observing outputs~\cite{papernot2017practical}. 
In addition, more work should study how query number and model complexity relate in practice. Such results would also hold implications for other inference attacks based on test queries like inversion attacks~\cite{jegorova2022survey} or model extraction~\cite{jegorova2022survey}.

\myparagraph{Membership and attribute inference} We found evidence of the applicability of membership and property inference threat models in practice. We should thus address the ongoing arms race to defend such threats~\cite{jegorova2022survey}.  
Within our sample, the threat model for attribute inference was rare (Sect.~\ref{sec:privacy}).
In membership, we recommend investigating attacks being staged without control over the submitted test queries (Sect.~\ref{sec:privacy}). For both attacks, more understanding of minimal knowledge attacks would be beneficial, for example, to infer membership for several points from only one query. 

\myparagraph{Security relevance of libraries} We found evidence that energy-saving libraries are frequently ($>$33\%) used in practice (Sect.~\ref{sec:sloth}). It would thus be beneficial to study sloth attacks, in particular with a focus on different energy-saving approaches, and whether attacks transfer across them or not. In addition, it is important to comply upfront with the corresponding constraints of practical training (Sect.~\ref{sec:trtimeatt}) and test-time (Sect.~\ref{sec:testtime}) threat models.

\myparagraph{Attack cost and stealthiness} 
There are possible costs attached to querying and changing data. A general focus on attacks with very few required resources is thus beneficial to understanding real-world vulnerabilities. Another aspect is the stealthiness of the altered data, which is already object to debate~\cite{gilmer2018motivating}. This stealthiness may also be related to domain knowledge (Sect.~\ref{sec:genattacksdefenses}), where more work is required to understand the nature of these constraints and how frequently they occur in practice. This is also loosely related to the question of whether and to what degree the attacker needs to know the exact data distribution of the victim. Finally, stealthiness needs to be studied in relation to human perception~\cite{elsayed2018adversarial}, but also in the context of the limitations of automated detection. 

\subsection{Practical Implications}
Our research has implications beyond AI security research, which we discuss now. \new{The most important implication is that our anonymous participants' AI systems may be vulnerable, highlighting the need for deployable, practical mitigations.}

\myparagraph{Regulatory and societal implications} 
Assessing the true vulnerability of AI systems in practice is required by legislative approaches and regulatory frameworks such as the EU AI Act. With poisoning and evasion, Article 15 of the latter even explicitly names some threat models whose underlying assumptions we could confirm (Sect.~\ref{sec:results}). Technical solutions and organizational measures to address such vulnerabilities are relevant. However, our study shows that the legal text’s addition ‘where appropriate’\footnote{\url{https://data.consilium.europa.eu/doc/document/ST-5662-2024-INIT/en/pdf}} is crucial. Not every threat model applies for each AI system as access schemes vary (Table~\ref{tab:threatModelInd}) and may be related to use-case, industry area, and other factors, which are left for future work. We also took a first step towards understanding what influences security-relevant features of models (Sect.~\ref{sec:threatmodelfactors}). More work is needed here as well. Yet, our results show that security is not a primary influence, implying that to make AI systems more secure, regulations are needed to prevent possible future incidents. Beyond regulation, assessing vulnerabilities helps to manage the risk of potential security incidents. Using our threat models (Sect.~\ref{sec:results}), the risk assessment of AI products in practice can now be completed as previously unknown settings can be studied. In this sense, our work has the potential to reduce what formerly were blind spots in AI systems.

\myparagraph{AI security in practice}  All 6 attacks studied within the framework of AI security are theoretically possible in practice. 
The low reported percentage of vulnerable settings reported within our sample is however rather small, potentially contributing to an explanation of few found AI security incidents~\cite{grosse2022so}. Furthermore, although common academic dataset specifications do occur in practice, many participants reported small feature sizes and large numbers of samples (Sect.~\ref{sec:datasets}). Understanding the effect of a small feature space with a potentially large amount of training is thus required. Analogously, we need to understand the limitations of not knowing data in practice. In tasks such as malware detection, feature encodings are secret, limiting the attacker~\cite{biggio2018wild}. More work is needed to understand these limitations and how frequent they are in practice. Orthogonally, we recommend more work studying what influences the exact configuration of threat models in organizational contexts (Sect.~\ref{sec:threatmodelfactors}). A deep understanding of which threat models are used in which cases could help to anticipate and mitigate vulnerabilities, but also understand which properties enable vulnerabilities in the first place. 

\section{Related Work}\label{sec:relwork}
While several contributions criticize existing AI security threat models~\cite{gilmer2018motivating,evtimov2020security,apruzzese2022real,cina2022energy}, to the best of our knowledge, no other work provides an overall picture of this research gap.

The closest related work to this study is a questionnaire-based quantitative study by Grosse et al.~\cite{grosse2022so}. They study AI practitioners’ AI security perception, asking about practitioners’ general concerns and individual attacks. Grosse et al. also used statistical tests to investigate influences on attack concerns. In our work, in contrast, we study whether academic threat models and AI usage in practice are aligned, and thus if AI security concerns are justifiable. 

Several works collect loosely similar information as us, including Kaggle's annual report about ML and data science~\cite{kaggle}, which provides information on, for example, the algorithms used in practice. Furthermore, Nahar et al.~\cite{nahar2022collaboration} investigated the origin of the used data in a small qualitative sample and the data engineer's effect on data requirements. Dilhara et al.~\cite{dilhara2021understanding} studied the usage of libraries in ML-based code within public repositories, not industry applications. Renieris et al.~\cite{renieris2023building} examined the practical usage of third-party tools and found that almost three-quarters use such tools. The same authors~\cite{renieris2023building} show that such tools may cause AI failures. Finally, Mink et al.~\cite{mink2023security} investigate why the number of deployed AI security mitigations in practice is rather low.  

Previous works reported low~\cite{boenisch2021never} to medium~\cite{grosse2022so}  AI security concern by industrial practitioners---our work indicates that this impression may stem from an indeed small attack surface due to little granted access to AI models in practice.

\section{Conclusion}\label{sec:conclusion}
We took a significant step towards more practical AI security research. We surveyed common threat model properties in practice 
and matched these to 6 threat models from AI security research. Our findings have implications for current legislative attempts like the EU AI Act that require security and vulnerability assessments of AI systems. We also set previously low numbers of AI security incidents into context, although our sample provides only initial insights on vulnerabilities through 3rd party accessible channels. Real-world vulnerabilities may be higher. Our work also
paves the way toward a deep understanding of the security of AI-based products in practice. Most importantly, while academia, despite criticism, has elaborated valid threat models, we also identify significant gaps. Current threat models are too generous about, for example, training data access or test time queries. More practical threat models should be researched. At the same time, attacks requiring few resources can potentially be applied to a large fraction of models. A black-box evasion attack with less than 100 queries, for example, could target \new{28.8-44.4\%} of the models in our sample. Even if just a small portion of companies match the exact constraints assumed in academic papers, these systems have to be defended.

\section*{Acknowledgments}
We thank our participants, Hyrum Anderson, Marielle Dado, Daryan Dehghanpisheh, Nikki Hogg, Ritesh Sharma, Aryan Trip, Karn Wong, Alla Zhdan, and the MLOps community 
for their support. We also thank the anonymous reviewers and our shepherd for their valuable feedback.

\bibliographystyle{plain}
\bibliography{lit.bib}
%


\theendnotes
We depict our study's full questionnaire.

\myparagraph{I - Demographics}

\question{1}: How old are you? 
[18-24, 25-34, 35-44, 45-54, 55-64, 64+]

\question{2}: What gender do you identify with?  
[Female, male, other, I do not want to disclose]

\question{3}: In which country are you located? Please use the country calling code of your country to choose a group, where X can be replaced by any digit. \\
$[+1, +299]$ (North America) \\
$[+2X]$ (Africa) \\
$[+30-+35X, +39]$ (Southwest Europe) \\
$[+36X-+38X]$ (East Europe) \\
$[+4X]$ (Central / Northern Europe) \\
$[+52, +53, +50X]$ (Central America) \\
$[+51, >+54]$ (South America) \\
$[+6]$ (Oceania, Australia, New Zealand) \\
$[+7, +976]$ (Russia, north Asia) \\
$[+8]$ (East Asia, Japan) \\
$[+90,+96X,+970-+974]$ (Near East and Türkiye) \\
$[ +91-+95, +99X, +975, +076, +977]$ (Southeast Asia) 

\question{4}: What is your level of education? Please specify the highest.  
[Highschool, Bachelor, Master / Diploma, Training / Apprenticeship, PhD, Other]

\emph{The four following questions (5-8) have all the same replies, namely:}
[None, Education only, $<$ 1 year, 1-2 years, 2-5 years, 5-10 years,  $>$ 10 years]

How many years have you worked in/with...

\question{5}: AI/ML? \hspace{1.5em} \question{6}: Security? \hspace{1.5em}
\question{7}: AI/ML Security?

\question{8}: AI/ML policies or AI/ML risk management?

\question{9}: In which country is your organization headquartered?
[replies as above in \question{3}]

\question{10}: What is the number of employees at your organization? 
[0-49, 50-249, 250-999, $>$1000]

\question{11}: What is the size of the team you work in?
[$<$3, 3-5, 6-9, 10-15, $>$15]

\question{12}: In which industry area does your company operate? 
[Mining, utilities, construction, manufacturing, wholesale trade, resale trade, transport. \& warehousing, information, finance \& insurance, real estate \& rental \& leasing, professional \& scientific \& tech. services, management of companies \& enterprises,  administration, education, health care \& social assistance, arts \& entertainment \& recreation, accommodation \& food services, public administration, other services]

\question{13}:  To encompass specific industries, please tick which of the following areas you work in (feel free to tick several)?  
[academic research, automotive or suppliers of automotive, cyber security, healthcare, none of these]

\myparagraph{Part II.A - Your AI based Projects (33\% of survey done)}

\question{14}: What is the status of the ML projects you work on?
[Indirect usage (e.g. certification, auditing); Evaluating use cases; Starting to develop models; Getting developed models into production, Models in production, for 1-2 years; M. in production, for 2-4 years; M. in production, for $>$5 years]

\textbf{In case you work on several projects, from here on, please stick to one (for example your favorite) AI project.}

\question{15}: Are you collaborating with domain experts on our data? As an example, consider a healthcare application where a doctor or domain expert needs to be involved.
[yes; no, not required;  no, none available] 

\question{16}: How clear is the specification (in terms of intended use, infrastructure, deployment, etc) of a requested model to you?
[linear scale from 0 (very unclear) to 100 (very clear)] 

\question{17}: How time-critical is obtaining a result for a query of your AI-based application?
[not time-critical at all; time-critical, but not real-time;  real-time required]

\question{18}: Are your inputs/features secret because they are customized or hand-engineered? For example, your company-developed representation of a program would be secret, whereas an RGB image encoding would be well-known. 
[linear scale from 0 (very secret) to 100 (well known)]

\myparagraph{Part II.B - Development of AI (45\% of survey done)}

\question{19}: Which of the following resources do you use to develop your models in terms of code? (several replies are fine)
[self written code; open source code; proprietary solutions]

\question{20}: Where do you train your models (several replies are fine)? 
[on-premise servers; cloud-provided servers; mix of both/hybrid cloud]   

\question{21}: Do you use pre-trained third-party models, in other words, models not trained on your own data? (several replies are fine)
[yes; yes, but we fine-tune them; yes, but for development only; not in deployment; no]  

\question{22}: Do you use any libraries, software, or self-written code to decrease runtime or execution of your models at deployment? Examples would be quantization, ASIC sparsity-based models, etc.
[yes;  sometimes; no]

\myparagraph{Part II.C - Data and Model within your AI-based Project (55\% of survey done)}

\emph{The four following questions (23-26), have all the same replies, namely:}
[Accessible to 3rd party; under access control; not accessible at all]

Please specify the accessibility of different parts of your ML pipeline to third parties.

\question{23} Training data. \hspace{2em} \question{24} Model (parameters).

\question{25} Test data. \hspace{3.8em} \question{26} Model outputs.

\textbf{For the next questions, educated guesses for the responses are sufficient.}

\question{27}: Which fraction of your test data comes from public sources (e.g., Internet)?
[None; $<$1\%;  1\% -5\%;  5\% -10\%;  10\% -15\%;   25\% -50\%;  50\% -75\%;  $>$75\%; I don’t know] 

\question{28}: Which fraction of your training data comes from public sources (e.g., Internet)?
[None; $<$1\%;  1\%-5\%;  5\%-10\%;  10\%-15\%;   25\%-50\%;  50\%-75\%;  $>$75\%; I don’t know] 

\question{29}: What is the size of your input (e.g., number of features)?
[$<$ 10; 10 - 100;  100-1K; o 1k-100K; 100k-110M; $>$100M;  does not apply]

\question{30}: How many samples do you train on?
[$<$10, 10-100;  100-1K; 1k-100K, 100k-100M, $>$100M;  does not apply]

\question{31}: How many samples do you evaluate your model on?
[$<$10, 10-100;  100-1K; 1k-100K, 100k-100M, $>$100M;  stream of data]

\question{32}: How many model outputs can a third party observe?
[$<$10, 10-100;  100-1K; 1k-100K, 100k-100M, $>$100M;  unconstrained]

\question{33}: How many model outputs can a third party query from your model during the entire time the model is available?
[$<$10, 10-100;  100-1K; 1k-100K, 100k-100M, $>$100M;  unconstrained]

\question{34} Are you able to estimate the performance (expected accuracy) before training? As an example, a very clear case is a known classification with documented accuracy $>$90\%.
[linear scale from 0 (not at all) to 100 (absolutely)] 

\question{35} How easy is it for you to assess the quality of your training data? As an example, is the data easy to inspect visually, or can you test whether it corresponds to your task?
[linear scale from 0 (not at all) to 100 (absolutely)]

\question{36} How much can you influence or enforce requirements (such as sampled from a certain source, inspected by a real worker, etc) on the used training data?
[linear scale from 0 (not at all) to 100 (absolutely)]

\myparagraph{III - AI security  (88\% of survey done)}

\question{37}: How relevant is the security of your AI-based product to you?
[1 (very low) to 100 (very high)] 

\question{38}: How relevant is the user's privacy of your AI-based product to you? 
[1 (very low) to 100 (very high)] 

\question{39}: How high do you estimate the risk of becoming a victim of an attack related to your AI-based workflows, products, or systems within the next 12 months? 
[1 (very low) to 100 (very high)] 

\question{40}: How likely do you estimate the probability of noticing an attack on your AI-based workflows? 
[1 (very low) to 100 (very high)] 

\question{41}-\question{43} as in Grosse et al.~\cite{grosse2022so}.

\question{41}: Did you already experience a circumvention of your AI-based workflows, products or systems? [yes/no] 

IF YES: 
\question{42}: How many circumventions of your AI-based workflows, products or systems have you experienced? [1,2,3,4,$>$4] 

\question{43}: Please describe the most severe circumvention of your AI-based workflows, products or systems. [text field] \\

\end{document}